# QEML: (Quantum Enhanced Machine Learning)
## Using Quantum Computation to implement a K-nearest Neighbors Algorithm in a Quantum Feature Space on Superconducting Processors

Siddharth Sharma

*Abstract*— Machine learning and quantum computing are two technologies which are causing a paradigm shift in the performance and behavior of certain algorithms, achieving previously unattainable results. Machine learning (kernel classification) has become ubiquitous as the forefront method for pattern recognition and has been shown to have numerous societal applications. While not yet fault-tolerant, Quantum computing is an entirely new method of computation due to its exploitation of quantum phenomena such as superposition and entanglement. While current machine learning classifiers like the Support Vector Machine are seeing gradual improvements in performance, there are still severe limitations on the efficiency and scalability of such algorithms due to a limited feature space which makes the kernel functions computationally expensive to estimate. By integrating quantum circuits into traditional Machine Learning, we may solve this problem through the use of a quantum feature space, a technique which improves existing Machine Learning algorithms through the use of parallelization and the reduction of the storage space from exponential to linear. This research expands on this concept of the Hilbert space and applies it for classical machine learning by implementing the quantum-enhanced version of the K-nearest neighbors' algorithm (an existing lazy-learning deterministic classifier). The primary experiment of this research is to build a noisy variational quantum circuit KNN (QKNN) which mimics the classification methods of a traditional K-nearest neighbors' classifier. The QKNN utilizes the distance metric of Hamming Distance and is able to outperform the existing KNN on a 10-dimensional Breast Cancer dataset.

I. INTRODUCTION

Quantum Computing is a new paradigm of algorithmic study which extends quantum mechanical phenomena to the world of traditional computing. In 1982, Richard Feynman proposed an initial quantum computer, which would have the capacity to facilitate traditional algorithms with quantum circuits [1]. To understand systems of electrons and to navigate the multiple independent probabilities of electron location based on quantum phenomena, Feynman envisioned the concept of a quantum computer; he believed that quantum computers could ideally simulate quantum behavior as it would have occurred in nature. The quantum systems which Feynman wished to simulate could not be modeled by even a massively parallel classical computer. For example, let us consider the probability calculations of multiple particle systems. If we have two electrons constrained to being at two points (A and B), then there are 4 possible probabilities of their location (both at A, one A – one B, one B – one A, both at B, etc.). For 3 electrons, there are 8 probabilities, for 10 electrons, there are 1,024 probabilities, and at 20 electrons, there are 1,048,576 probabilities. Therefore, it is easy to see that measurements get out of hand for traditional physical systems with millions of electrons. Quantum computation can efficiently solve this problem.

On the other hand, Machine Learning has been able to solve conventional problems through innovative techniques on classical computing devices. Machine learning is a data-dependent technique and involves employing pattern recognition for set-sized datasets. However, Machine learning algorithms for both regression and classification see clear performance drops as the size and number of features for the given problem grow. Also known as the *Curse of Dimensionality*, this problem has long plagued machine learning algorithms and has led to polynomial growth and drastic increase in runtimes. The root of this problem lies within the storage of data itself. For classical machine learning, the states and properties of certain vectors are placed within a classical feature space. This feature space is the reason that the kernel functions and performance of certain machine learning algorithms face exponential storage and runtime. In recent research, it has been proposed to implement quantum computing to reduce this burden of exponential storage, thus creating a quantum feature space. The direct solution is to transform the classical states into quantum states that are able to be stored more efficiently. Several quantum machine learning algorithms have already been proposed and implemented – qSVM, qVC, etc.

The goal of this paper is to implement the quantum variant of the classic K-nearest neighbors classifier through the metric of Hamming distance in a quantum feature space. We will begin by examining the methods of a K-nearest neighbors and understanding how a quantum-enhanced feature space would work. We will then proclaim and identify the key properties of Quantum Machine Learning as a whole and the intuition for why it should improve the performance of the classic KNN algorithm. The next step is to justify and implement basic properties and algorithms that form the backbone of a Quantum KNN, namely Fidelity and Grover's algorithm. We will then examine the techniques used to create the Quantum SVM and justify the use of Hamming distance as a distance measure. It will be shown that the distance in a quantum feature space can be calculated efficiently on a superconducting processor (IBMQ simulator). We will then explain the mathematical basis of a QKNN and explain the various gates and subroutines that will need to be used to implement it. Finally, a standard ad-hoc Breast Cancer dataset will be used to qualify the performance of the Quantum K-nearest neighbors against its classical counterpart. This method will be considered alongside the necessary post-hoc analysis and considerations of the benefits of a quantum enhanced classifier.

## II. K-Nearest Neighbors

The K-Nearest Neighbors (KNN) algorithm is a simple example of supervised learning (input being mapped to output). KNN is a classification algorithm and classifies an input as a discrete or categorical output. Other examples of supervised learning classifiers include decision trees, naïve Bayes, and random forest models. Since KNN is an instance-based lazy learning algorithm, it is usually simple and intuitive to train. The fundamental assumption for the KNN algorithm is that datapoints with similar behavior exist in close proximity to each other. If this assumption does not remain fulfilled for a certain sample problem or dataset, the KNN model will not provide statistically significant results. In other words, KNN captures the concept of "closeness" or proximity. Thus, the KNN algorithm is successful in sample datasets where similar outputs cluster together in close proximity. The algorithm works by computing the distances from a specified example to other local examples. The algorithm is known as "K"-Nearest Neighbors since K nearby points are examined. K is a parameter which can be chosen and tuned via iterations of training a KNN algorithm. In most cases, K is an initial arbitrary point. Pseudocode for KNN is given below.

### I. KNN Algorithm (1)

```
Algorithm 1 KNN algorithm
Input: x, S, d
Output: class of x
for (x', l') ∈ S do
    Compute the distance d(x', x)
end for
Sort the |S| distances by increasing order
Count the number of occurrences of each class l_j
    among the k nearest neighbors
Assign to x the most frequent class
```

Since the KNN algorithm is based on a distance metric (typically Euclidean distance), we must evaluate this for two sample feature vectors $(\tilde{v}, \overline{w})$. The goal is to identify the class of a new feature vector $(\overline{u})$ based on its distance to the two other feature vectors. The vector, $\overline{u}$, is assigned the class of the feature vector that it was closer to. The two parameters which affect the performance of KNN the most are the value of K and the number of dimensions $(n)$. A smaller value of K could allow noise to have a major influence on the prediction (majority vote), whereas a large value of K makes the problem more computationally expensive. The common consensus is to give K the value of $\sqrt{N}$ where $N$ is the number of training data points. Thus, the value of K heavily affects performance of the classifier.

On the other hand, it is important to note that a KNN classifier suffers from the "*Curse of Dimensionality*". This is best illustrated in an example. If we have 1,000 training data points uniformly distributed across our vector space and our test data point is at the origin, the performance rapidly varies with dimension. In 1-D space, it takes about a distance of 5/1,000 = 0.005 on average to get 5 nearest neighbors. In 2-dimensional space, it takes about a distance of $(0.005)^{1/2}$ and in $n$-dimensional space, it takes about a distance of $(0.005)^{1/n}$ in each direction as the training data points become sparsely distributed when the dimension of the space increases. The goal of a Quantum KNN (QKNN) approach would be to compute the distance via a quantum algorithm that is both efficient and simple to scale while minimizing both CPU time and cost of data point storage.

## II. Quantum-Enhanced machine learning

*Quantum-Enhanced Machine Learning* (QEML) involves taking supervised learning algorithms and making them more efficient through the use of quantum gates and orthogonal transformations to achieve more meaningful results. QEML strives to also offer solutions to the challenges of both data storage and slower execution. Before constructing a Quantum K-Nearest Neighbors Algorithm, it is necessary to explore the limitations of previous quantum algorithms while understanding the advantageous nature of quantum supervised learning. The primary properties of QEML can be expressed in the following aspects: improved representation space and acceleration of algorithm execution due to use of quantum heuristics. The storage space can be exponentially reduced through the use of *quantum superposition*. As discussed earlier, superposition is a fundamental quantum property which allows qubits to hold multiple states at once. For example, a $N$ qubit state $|\phi_1, \phi_2, ..., \phi_3\rangle$ can also be written as:

$$|\phi_1 \phi_2 \cdots \phi_n\rangle = \sum_{i=0}^{2^n-1} c_i |i\rangle \quad (2)$$

In a quantum computer, all binary numbers in the set $\{0, 1, ..., 2^{n-1}\}$ exist in an $N$ qubit quantum register. In a classic computer, only 1 binary number in the set $\{0, 1, ..., 2^{n-1}\}$ can be stored in an $N$ bit register. This example shows the benefit that quantum computers can provide in terms of reducing the size of the feature space (storage scale). Quantum computers achieve this through the aforementioned property of superposition which allows a $N$ qubit register to hold exponentially more binary numbers without issues in size and scale. Thus, this property promotes the idea of an *enhanced feature space* in QEML classifiers.

The second benefit that quantum computing offers to machine learning is that of acceleration during execution of machine learning algorithms. This property of QEML is also known as "*quantum parallelism*". "Quantum parallelism" similarly arises from the ability of a quantum register to exist in a state of superposition. Each component of superposition can be represented as a function. Each component of the superposition is evaluated by its respective function within the quantum register. Since the number of possible states is $2^n$ where $n$ represents the number of qubits, it would take a classical computer an exponential number of operations to perform a task. However, a quantum computer can mitigate this issue through superposition and quantum parallelism, performing a similar task in one operation. The following example mathematically demonstrates this effect. Assuming a quantum environment, if a unitary operator $U_f$ is transformed by the function $f(x)$, $U_f$ must accomplish the task by inputting $x$ from $|00 \ldots 0\rangle$ to $|11 \ldots 1\rangle$. For a classic computing environment, computing $f(x)$ from $x$ inputs would take $2^n$ cycles or would require $2^n$ CPUs working in parallel. The relationship between $U_f, f(x)$ and the $x$ inputs is outlined in the equation below:

$$U_f\left(\frac{1}{\sqrt{2^n}}\sum_{x=0}^{2^n-1}|x\rangle\right) = \frac{1}{\sqrt{2^n}}\sum_{x=0}^{2^n-1}U_f|x\rangle = \frac{1}{\sqrt{2^n}}\sum_{x=0}^{2^n-1}|f(x)\rangle \quad (3)$$

In summary, quantum algorithms offer an elegant solution to problems faced in classical learning since quantum algorithms can store all training data points of exponentially large size as a linear size due to superposition; due to entanglement and interference, they can also compute distances near-simultaneously.

III. FIDELTY AND DISTANCE COMPUTATION

Even with the benefit of quantum parallelization, retrieving information from a quantum state with high performance is a difficult task. This is because during the process of measurement of a quantum state, there tends to be partial collapse of the quantum state and loss of previous information attained by quantum algorithms. To get the computing function $f(x)$ requires an innovative approach. Buhrmann's technique (quantum fingerprinting) is a simple solution to the problem of loss of quantum state and helps to calculate the distance of two vectors with high accuracy and fast execution. His process is detailed as the following: The auxiliary qubit $|0\rangle$ is first transformed through the left Hadamard gate to $\frac{|0\rangle+|1\rangle}{\sqrt{2}}$. The circuit then employs a **SWAP** gate to switch the two vectors $|x\rangle$ and $|y\rangle$ (i.e. $|xy\rangle \to |yx\rangle$). This process has also been illustrated in the diagram below:

II. SWAP Circuit (4)

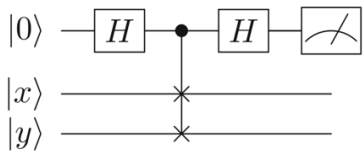

The above quantum circuit is also known as a Controlled Swap ($C - SWAP$) test and is essential to QEML since it provides for the property of *fidelity* (analogous to *cosine similarity* in classical machine learning). Fidelity measures the similarity of two quantum states ($|\phi\rangle, |\psi\rangle$). Fidelity can be represented as $|\langle x|y\rangle|$. If the two quantum states are orthogonal, the fidelity is 0; when the quantum states are identical, the fidelity is at its maximum which is 1. Another important note about fidelity concerns the efficiency of the quantum solution; when the dimension of the quantum state vectors is higher, the quantum solution that uses fidelity becomes more efficient. For developing a quantum analog to the K-Nearest Neighbors algorithm, a quantum solution for the calculation of distance must be carried out. In classical machine learning, the distance between labeled examples in the vector space is typically calculated by the *Euclidean Distance* ($\sqrt{\sum_{i=1}^{n}(q_i - p_i)^2}$). Using the earlier discussed trick of fidelity, we can represent the concept of Euclidean Distance in a quantum space: $\sqrt{2 - 2|\langle x|y\rangle|}$ .

Since most supervised machine learning classifiers are based on concepts of class similarity and distance measurement, this trick (representation of distance via fidelity in a quantum space) is foundational to the performance of QEML algorithms. Some examples of algorithms and metrics used for computation of the distance in Quantum Machine Learning include Grover's Algorithm, the Hamming Distance, Lloyd's Algorithm, and Schor's Algorithm. These are all closely related to *quantum amplitude estimation* and attempt to measure distances between labeled examples in a quantum space. Another key breakthrough in quantum amplitude estimation is seen in the development of the quantum minimum search algorithm (similar to QESA).

IV. INITIAL SIMULATION OF FIDELITY WITH QISKIT

To build a Quantum K-Nearest Neighbors algorithm and to compare it to its machine learning counterpart, the distance will be measured via the concept of fidelity and the Controlled Swap gate. For measuring distance with a metric, Hamming Distance is optimal for discrete values whereas Euclidean distance is a strong indicator for continuous values. To conduct an initial simulation of fidelity using a quantum computer, we must use a python library known as *Qiskit*. Qiskit is a popular framework which allows users to simulate quantum circuits on a classical computer. To designate the earlier defined ($C - SWAP$) gate, we must first build the circuit in the IBM Quantum Experience simulator. Since Qiskit is an open-source framework which provides access to building circuits on noisy quantum computers, this is not a challenging task.

While its main goal is to facilitate quantum research in open areas in quantum computing, Qiskit also consists of four smaller libraries which allow developers to build full-stack quantum circuits: *Aqua*, *Aer*, *Terra*, and *Ignis*. Aer is designed to accelerate development through use of simulators and debuggers. Aqua is for building algorithms and larger quantum circuits. Terra is the library which propagates the code foundation of quantum circuits. Lastly, Ignis addresses issues with noise and interference in quantum circuits. To build a circuit to prove the concept of fidelity, we will use Aer and Terra in conjunction to both build the quantum circuit and to analyze results and produce the state-vectors.

## V. RESULTS OF SIMULATION WITH IBMQ

First, we build the variational quantum circuit and its respective quantum registers within the IBM quantum experience platform. To illustrate fidelity and to build the $C-SWAP$ gate, we first instantiate our register with 3 qubits ($q_0, q_1, q_2$) and 3 bits ($c_0, c_1, c_2$). The complete quantum circuit is shown below:

III. IBMQ Simulation (5)

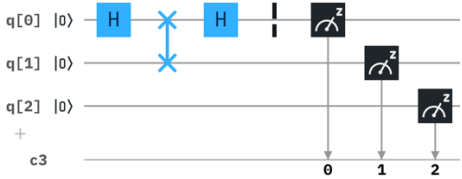

To transfer our first qubit ($q_0$) into a state of superposition, we first apply a Hadamard ($H$) gate. The other two qubit registers ($q_1, q_2$) store the transformed state of the first qubit. We then apply a Fredkin gate (the gate in-between) to conduct the $C-SWAP$ operation. The final signal passes through the second Hadamard gate and is then measured via the Z-measurement gate. If the initial states were orthogonal, the probability is 0.5, while if they were identical, the probability is then 1. To physically simulate the circuit, two backends were used. The *state-vector* and *BasicAer* (unitary) backends were used. The unitary result was a matrix of all possible state-vectors. If the state-vectors are roughly equally distributed (after noise) in a $C-SWAP$ circuit, then we have reaffirmed the concept of fidelity. The results from the aforementioned circuit were promising when simulated with the backends in the IBMQ Experience. The unitary state-vectors are illustrated below:

IV. Unitary State-Vectors (6)

```
[ 0.5+0j, 0.5+0j, 0.5+0j, 0.5+0j,
  0+0j, 0+0j, 0+0j, 0+0j ]
```

This unitary state-vector reaffirms the concept of fidelity since the value is usually 0.5 (both the qubits are orthogonal) or 0. This is very similar to cosine similarity and reflects the instrumental trick which was discussed earlier. Now that fidelity has been illustrated, it may be employed to calculate the distance. Since the main goal of this paper is to build a quantum variant of the K-nearest neighbors, this simulation is a promising start to calculating distance in the QKNN model.

## VI. ANALYSIS OF QUANTUM SVM

Prior to designing the QKNN, it is important to understand the development of quantum kernel methods and the implementation of the enhanced quantum feature space, one of the central tenets of QEML. We will now analyze the quantum variant of the Support Vector Machine (QSVM), another popular kernel method for classification. The benefit of QSVM is that it provides for an enhanced feature space and thus, improved performance and storage capabilities. The classic SVM is known as a kernel method, a ubiquitous technique in pattern recognition. As described in the paper [1] by Havlicek et al., classical information is mapped to a quantum state. For background, we are given data from a training set $T$ and a test set $S$ of a subset. Both are assumed to be ground truth labeled by a map $m: T \cup S \to \{+1, -1\}$ unknown to the algorithm. The training algorithm only receives the labels of the training data $T$. The goal is to infer an approximate map on the test set S, where we are able to map the output from a given input to the set, $\{+1, -1\}$. This output should ideally closely correlate with the ground truth labeled map ($T \cup S \to \{+1, -1\}$). In a classical Support Vector Machine (SVM), the data is mapped non-linearly to a higher dimensional space where it is later separated into distinctive clusters via a defined segmentation, also known as a *hyperplane*. Havlicek et al. were able to generate a defining hyperplane in the quantum feature space as expressed in the following: $\Phi: \vec{x} \in \Omega \to ||\Phi(\vec{x})\rangle \ |\Phi(\vec{x})\rangle|$. Their mapping of a hyperplane in the quantum feature space is modeled below:

V. Havlicek Hyperplane (7)

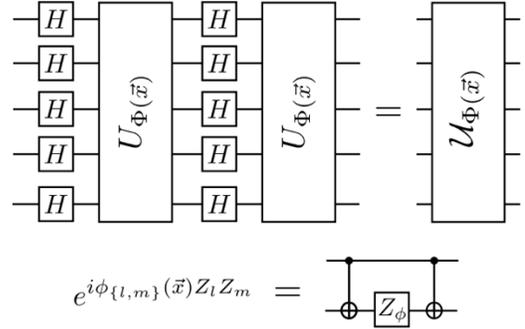

Havlicek et al.'s second goal was to implement a variational circuit which is normally tremendously difficult to operate on a classical computing system. To be able to design the QSVM, they initially defined the feature map on *n*-qubits generated by the following unitary ($\mathcal{U}_{\Phi(\vec{x})} = U_{\Phi(\vec{x})} H^{\otimes n} U_{\Phi(\vec{x})} H^{\otimes n}$). They then generate the hyperplane through the aforementioned algorithm and equation below:

$$U_{\Phi(\vec{x})} = \exp\left(i \sum_{S \subseteq [n]} \phi_S(\vec{x}) \prod_{i \in S} Z_i\right) \quad (8)$$

Following the generation of the hyperplane to minimize the margin, they then optimize the algorithm for the parameters ($\vec{\theta}, b$) in the noisy experimental setting via *Spall's SPSA* (Stochastic Gradient Descent Algorithm). Havlicek et al.'s final variational quantum classifier $W(\vec{\theta})$ is able to perform both binary and multiclass classification with high accuracy as exemplified in their experiments with a 5-qubit processor. The efficacy of their results (a successful variational quantum classifier that is able exploit quantum feature space) improves with dimension *n*. By analyzing their adoption of quantum-supported kernel methods like the QSVM, we gain useful intuition on the process behind building a QEML algorithm like the QKNN. This process has also been simplified due to pre-implementation of practical quantum algorithms on the Qiskit Aqua library. Some practical algorithms already available in Aqua include Grover's algorithm, quadratic estimator, the Fourier transform, eigen_solver, etc.

## VII. BUILDING A QKNN AND COMPUTING HAMMING DISTANCE

We will now lay down the foundation for distance calculation based on the earlier discussion of Schuld's trick and Havlicek et al.'s use of an enhanced quantum feature space. Based on analysis of metrics of quantum amplitude estimation, the Hamming Distance was found to be most optimal for higher dimension categorical learning problems. Hamming Distance was selected since it solves the "*Curse of Dimensionality*" for typical KNN. In a KNN algorithm, we usually have $N$ of $D$ patterns and access to the training set. The complexity of rating of one neighbor is $O(D)$. To rate all neighbors, we achieve complexity $O(ND)$. To find an additional "K" nearest neighbors on top of the already defined examples, we have complexity $O(KN)$. Thus, our total algorithmic complexity for KNN is shown to be $O(ND + KN)$. Thus, despite current technology and polynomial factorization, it is still a computationally expensive task to run the KNN algorithm for large datasets (especially in higher dimensions). Thus, we can solve this problem by adopting a quantized metric for computing distance (rating) between sample data points. This paper uses the *Hamming Distance* as the primary metric for assessing the distance between sequential data in a quantum vector space. We may now define the Hamming Distance: it is the number of positions at which the corresponding symbols of two-bit vectors of equal length are different. The operator for a Hamming distance is $\leftrightarrow$. This is best illustrated with a few examples: 00101↔00101 = 0, 00101↔00111 = 1, 00101↔10111 = 2. The Hamming distance metric always meets 3 criteria: 1.) non-negative, 2.) symmetric in nature, 3.) satisfies the triangle inequality. It is not immediately obvious how the Hamming distance is useful for determining the distance between two feature vectors. However, Hamming distance is used for a variety of practical applications (ex. text classification, image classification). Even a simple QKNN algorithm in Hamming space is competitive with the highest performing discriminative models. Hamming distance also allows us to skip time-consuming operations in manipulating quantum state such as phase estimation and tomography.

## VIII. SIMULATING GROVER'S ALGORITHM

The next algorithm which can be used to model a QKNN is Grover's Algorithm. Now that we have identified that we will find the distances between labeled examples in a quantum feature space using Hamming Distance, we can now use Grover's Algorithm to find the minimum spanning distance (find the nearest neighbors). Grover's Algorithm is illustrated below:

VI. Grover's Algorithm (9)

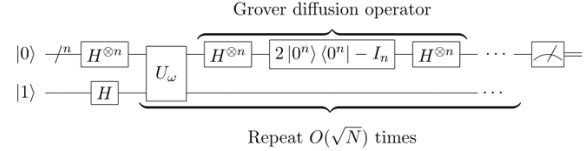

Grover's Algorithm is one of the best examples to demonstrate the physical speed-up that quantum computers provide. If we are searching through a dataset of dimension *n*, the time it takes to locate a certain data point takes $O(N)$ on a classical computer. However, if we use Grover's Algorithm and a quantum computer, this operation only takes $O(\sqrt{N})$ time in terms of its complexity. Grover's Algorithm is also known as an Oracle algorithm and it applies both unitary operators and amplify operator probability amplitudes (together, these denote the *Grover Diffusion Operator*). We can use the IBM Quantum Experience platform and Qiskit library to model the Grover Algorithm. An example of Grover's Algorithm for 2 qubits in a closed quantum space is given below:

VII. Grover's Algorithm Simulation (10)

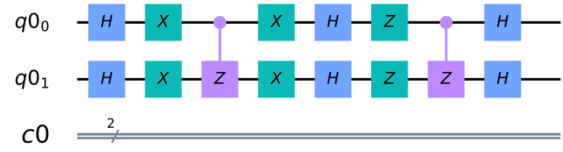

We have thus demonstrated that is both feasible and practical to implement Grover's Algorithm for the purpose of calculating the distance between points in a quantum feature map.

## IX. APPLICATION TOWARDS QKNN

We can summarize the background research in two approaches: the implicit and explicit techniques for building a quantum classifier. In the implicit approach, the kernel (quantum feature space) can be evaluated using traditional computation and can estimate the inner products via $\Phi: \vec{x} \in \Omega \rightarrow ||\Phi(\vec{x})\rangle \quad |\Phi(\vec{x})\rangle|$. The explicit approach involves finding the physical distances between data points in the "feature Hilbert space" of the quantum system. Techniques like the Hamming Distance and Grover's Algorithm allow us to conduct the explicit approach for quantum classification.

In an explicit approach, the model is solely trained by the quantum computer (no outside computation) and is able to output the prediction directly without the need for intermediate computation of the kernel. This approach corresponds with the recognized goals of QEML. It promotes both faster execution due to quantum algorithms and more storage capabilities due to an enhanced feature space in a quantum register. One technique to create a desirable quantum Hilbert space is through squeezing as proposed by Schuld et al. with their example of the Fock space. We may now summarize the overall explicit

approach for an image classification QKNN classifier in the following flow-chart:

VIII. Image QKNN Classifier (11)

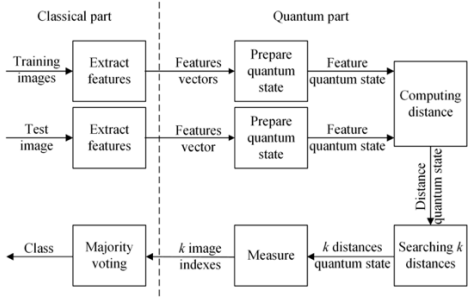

X. VARIATIONAL CIRCUIT FOR QKNN

We may now implement a QKNN through the aforementioned techniques and through Schuld et al.'s trick. We begin by instantiating each data point in our algorithm as a mathematical vector. All the vectors have $N$ features encoded and thus, the overall dataset can be represented as lying in an $N$-vector space. Classic supervised learning KNN algorithms seek to calculate distance to the K-nearest neighbors and do majority voting based on the class of the other nearest neighbors classes. In this case, we can use Hamming distance to approximate the distances to other neighbors. This solves the Curse of Dimensionality since it prevents the slowing on larger datasets due to repeated distance calculation. We begin by introducing the *Swap Test* for calculation of Hamming Distance. If we have the following state of an ancillary and an overlap:

$$|0\rangle \otimes |\psi\rangle \otimes |\varphi\rangle$$

We can apply two Hadamard gates with a Fredkin gate to provide the property of Fidelity:

$$\langle \Psi | (|0\rangle \langle 0| \otimes \mathbf{I} \otimes \mathbf{I}) | \Psi \rangle = \frac{1}{2} + \frac{1}{2} |\langle \psi | \varphi \rangle|^2$$

We may now begin to build the core module for calculation of Hamming Distance and quantum minimum search of the nearest neighbors. To begin, we instantiate our basic properties of a quantum feature space. For any QEML algorithm, if we map the features of our data to the ground quantum states in Hilbert space, it then becomes easier to select the $K$ nearest neighbors. The goal is for the final variational circuit to employ Schuld's trick of using fidelity whilst calculating Hamming distance for each vector, while also manipulating quantum parallelism with a $N$-qubit register. The core module for the above circuit was suggested by Kaye and is illustrated below:

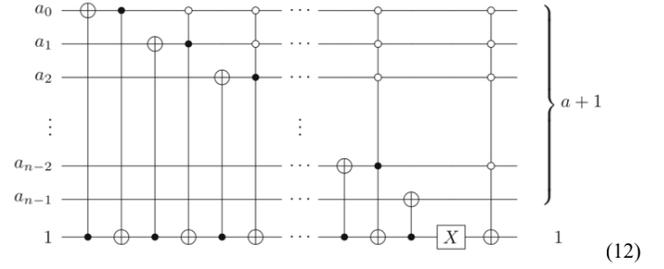

(12)

This circuit by Kaye is also known as the *Quantum a + 1 circuit* and uses incrementation. Prior to understanding this circuit, we must describe the setup. All bit vectors are mapped to their quantum ground state. ($0 \rightarrow |0\rangle$ and $1 \rightarrow |1\rangle$). The training set with $N$ feature vectors is represented in the following training set superposition:

$$|\mathcal{T}\rangle = \frac{1}{\sqrt{N}} \sum_p |v_1^p \ldots v_n^p, c^p\rangle$$

(13)

This circuit's underlying principle is to use addition between ancillary qubits. If a qubit ($a_0$) is "flipped" by this circuit, the circuit switches to the next least significant qubit. (i.e. $a + 1$). The number $a$ is within the bounds $[0, n-1]$. If a qubit $a[i]$ is "flipped" from 1 to 0, the addition continues. On the other hand, if $a[i]$ is flipped from 0 to 1, the addition stops and $a[i]$ is reset to 1, prompting the circuit to then continue. The workflow was demonstrated by Ran et al. [17] and is summarized below:

```
i = 0;
Do
    if a[i]==1 then {
        a[i]: 1 → 0;
        i++;
    }
    else
        a[i]: 0 → 1;
Until (a[i]: 0 → 1)
```
(14) from Ran et al.

The physical calculation of the Hamming Distance is now undertaken by this circuit and we must expand from the Quantum $a + 1$ circuit to a "$a + d$" circuit. We first apply a CNOT gate (to overwrite the first entry $a$ as 0 if $a$ = b) and later an X gate (to reverse the value). We now record the distances between all training points ($|x_1, \ldots, x_n\rangle$) and the new examples in the training set, ($|v_1, \ldots, v_n\rangle$)). We store the vector of computed distances ($|d^p_1, \ldots, d^p_n\rangle$). Since we had already constructed the quantum state ($|\phi_0\rangle$) in the first register, the training set ($|\mathcal{T}\rangle$) in the second register, and an ancillary qubit in the last register ($|0\rangle$), we can represent the labeling of neighbors (modification of the ancillary qubit $|0\rangle \rightarrow |1\rangle$). in the following unitary operation:

$$|\phi_3\rangle = U|\phi_2\rangle = \frac{1}{\sqrt{N}} \left( \sum_{p \in \Omega} |d_1^p \ldots d_n^p; v_1^p \ldots v_n^p, c^p; 1\rangle + \sum_{p \notin \Omega} |d_1^p \ldots d_n^p; v_1^p \ldots v_n^p, c^p; 0\rangle \right)$$

(15) from Ran et al.

The Hamming distances can then be represented in the summation $\sum_i d^p{}_i$ and can then be introduced into the previous Quantum $a + 1$ circuit to achieve an "$a + d_i$" quantum circuit. This new circuit is modeled below:

IX. $a + d_i$ quantum circuit

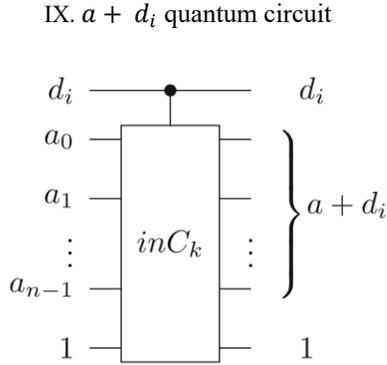

(16 a.) from Ran et al.

The quantum **OR** gate is now applied on top of the $a + d_i$ to achieve the final QKNN circuit. The **OR** gate was examined by Ran et al. [17] and is considered a sub-routine circuit which describes the condition of Hamming distance being less than $t$ qubits (where $a = l + t$). It is modeled below:

X. Sub-routine Circuit with OR Gate

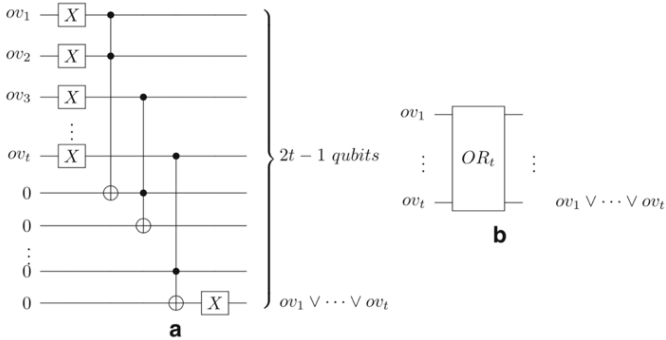

(16 b.), from Ran et al.

The majority voting in the classic KNN is best modeled in the QKNN as the probability of getting the final classification result. We can simulate this process through superconducting quantum processors. The IBM Quantum Experience platform allows us to simulate the QKNN circuit with quantum gates like **OR**, **ID** (null gate), **H**, **X**, etc. IBM Q works hand-in-hand with Qiskit (a Python module) to support quantum circuits. It is known that the computers built by IBM are built with superconducting transmon qubits and Josephson junctions to simulate a quantum environment. The issue with these machines is that experimental error is too difficult to control, and the perturbations and noise interferes in discriminative testing can make it tough to execute the algorithm. Thus, we may use a superconducting processor to analyze the results.

Drawing from the intuition of the "$a + d_i$" quantum circuit and Ran et al.'s work [14], we can implement a novel version of this Quantum K-nearest neighbors in IBM Q. We first implement the property of Fidelity and *Swap Test* to begin our circuit:

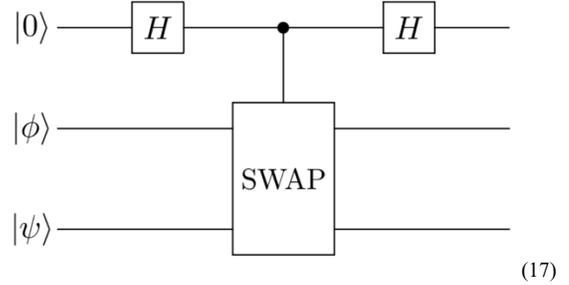

(17)

Simply put, the *Swap Test* allows us to measure the difference of the states of the vectors at the beginning of our circuit and integrates Fidelity (Schuld's trick). After instantiating two Hadamard gates at the beginning and conducting a **SWAP**, we see that the overall pattern for the rest of the circuit is a concurrent implementation of the "$a + d_i$" quantum circuit for $\log t$ qubits alongside the **OR** gate. The "$a + d_i$" quantum circuit uses Pauli X gates to apply a transformation that collectively finds the difference in vectors (Hamming distance). To account for gaps between the respective gates during calculation of Hamming distance prior to majority voting, we use **ID** gates (null gates). To implement the **OR** gate in the IBM Q, we must use a subroutine, a simplified circuit used in computation. The final circuit in IBM Q for 9 qubits is shown below:

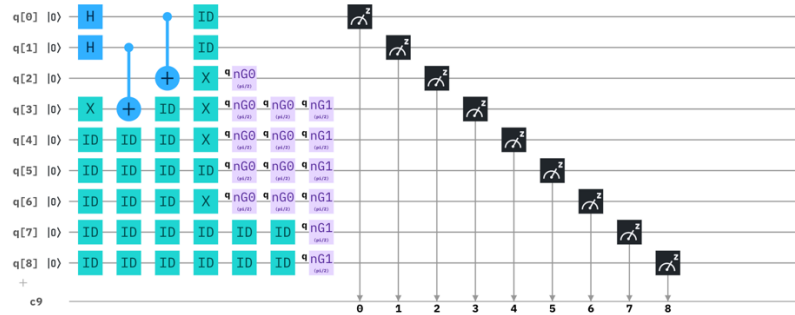

(18)

Since the main goal of this paper was to offer a side-by-side comparison of both classic ML and QEML, it would be incomplete without running tests to compare both the traditional KNN to its quantum variant (QKNN). The previously discussed benefits of QEML can be restated here: 1.) more storage capabilities and smaller solution space (due to enhanced feature space via superposition) 2.) faster execution of quantum algorithms due to quantum parallelization. We can now put the QKNN algorithm to the test by comparing its performance on a sample dataset against the classic lazy-learning KNN machine learning classifier. The dataset chosen for this problem was the *Wisconsin Breast Cancer Dataset* since it has $n = 10$ dimensions (multivariate) and is primarily useful for binary classification. 10 dimensions is considered optimal for gauging the performance of QEML against traditional machine learning since it is enough dimensions such that QEML should provide a tangible benefit but is not too large in that both ML and QEML will not become computationally expensive. We can first implement a KNN algorithm for the classification of tumors as either malignant (1) or benign (0). The dataset has the following attributes: diagnosis, radius_mean, texture_mean, perimeter_mean, area_mean, smoothness_mean, compactness_mean, concavity_mean, concave points_mean, symmetry_mean, etc. We can implement the KNN algorithm in a Jupyter Notebook. The first step is to input data as a *.csv* file and to analyze it for any missing cases or attributes. We then split into features and labels ($X$ and $y$). The data can be summarized in the following graph:

XI. Data Distribution

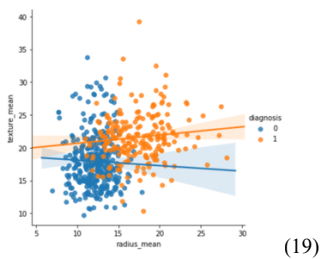

(19)

We can then use sklearn modules like model_selection.train_test_split to split the data into a reasonable train-test split (0.65 to 0.35 in this case). We can then use the Scikit-Learn KNeighborsClassifier class to fit a pre-made KNN algorithm to the breast cancer dataset. This specific KNN algorithm utilizes Euclidean Distance as its distance calculation metric and is thus comparable to the quantum approach. This version of the KNN algorithm is also unique in that it is designed solely for discrete classification. The inputs for classes for the algorithm were *NumPy* arrays and they were split using the aforementioned ratio. A nearest Centroid classifier was considered but was not substituted in place of a K neighbors classifier since there were no major benefits in the Centroid approach. The chosen KNN algorithm is also special in that it does not require traditional hyperparameter tuning. Rather, we must adjust the value for $K$ from its initial arbitrary value (usually 3). In this case, the optimal K (that produces minimal loss and misclassification) was determined to be $K = 13$ as seen below:

XII. KNN optimization

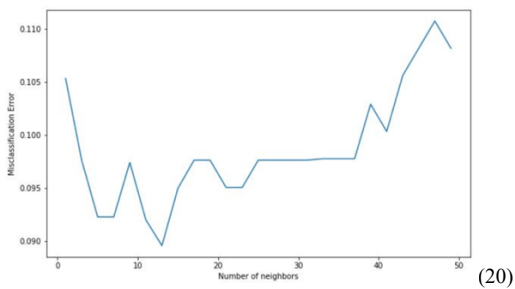

(20)

The final performance for the KNN algorithm was evaluated via the accuracy score metric of Scikit-Learn (average accuracy for the data with a certain number of neighbors during a training run). The QKNN circuit (fig. 22) was also used to do training on the IBM QASM simulator, providing similar results for accuracy. As earlier mentioned, this QKNN algorithm uses Hamming distance as the metric for calculating distance between neighbors in a quantum feature space. The final mean result for average performance/accuracy is shown below:

| Algorithm | Performance |
|---|---|
| QKNN | 0.9718 |
| KNN | 0.9627 |

These numbers come close to validating the original theory for why QKNN would outperform KNN in both performance and execution time due to quantum superposition. Although these results do not prove that QKNN has a statistically significant performance boost over KNN in terms of accuracy/performance on the same data, they provide a foundation for testing of quantum algorithms on much larger datasets. For example, one question that may arise is how is the performance difference between KNN and QKNN affected when there are $n = 100$ dimensions or $n = 1000$ dimensions? Answering these queries will require future analysis.

XI. CONCLUSIONS

Although this paper has proved that quantum computing applied to machine learning (QEML) does provide some intermediate benefit, we cannot definitively argue that QEML is superior to machine learning and will be the future of big data computation. These results were somewhat randomized and were computed on a limited dataset of $n = 10$ dimensions, so the conclusions can only be taken with a grain of salt. However, we were definitively able to simulate both the property of fidelity (analogous to cosine similarity in classic machine learning) and the path of Grover's algorithm for 2 qubits (analogous to minimum path search in classic machine learning). Despite the fact that QKNN could provide slightly more accurate results in a fraction of the time, the results are not statistically significant and that means that QEML has only been partially validated.

These experiments to simulate fidelity, the Grover's algorithm, and a QKNN were only possible due to open-source frameworks like the Qiskit python library and the IBM Quantum Experience platform. In the future when quantum computing is more accessible, we may have to conduct a cost-analysis to conclude whether the benefit is worth-it or not. Currently, physical quantum computers are not accessible to the wide population and are only accessible for researchers and collaborators at recognized institutions. Another important issue to recognize is that current quantum computers are quite susceptible to noise and physical perturbations during calculations. Moreover, decoherence is another key issue since loss/collapse of quantum information is still a common occurrence for today's quantum computers, and this will not be practical if physical quantum computing will one day be accessible to the majority of researchers. The ongoing "quest" to build a fault-tolerant quantum computer is still under development and will continue to require many more layers of abstraction and multiple paradigm shifts in the relationship between software and hardware. However, the success of the above experiments is a positive step for the field of quantum computing since it represents the synchronization of machine learning in quantum computing to benefit society as a whole. To summarize, quantum computing can benefit machine learning due to the idea of an enhanced quantum feature space. In essence, while fault-tolerant quantum computers are still decades away, we can still harness the power of quantum computing to improve both the efficiency and scalability of machine learning algorithms.

## XII. REFERENCES (IN-TEXT)

## XIII. ACKNOWLEDGEMENTS